# A high repetition rate picosecond LiNbO$_3$ THz parametric amplifier and the parametric gain study

Yuting Yang(杨雨婷), Gang Zhao(赵刚), Yue Huang(黄月), Liwen Feng(冯立文), Senlin Huang*(黄森林), and Kexin Liu(刘克新)

*State Key Laboratory of Nuclear Physics and Technology and Institute of Heavy Ion Physics, School of Physics, Peking University, Beijing 100871, China*

*Corresponding author: huangsl@pku.edu.cn

A high repetition rate, picosecond THz parametric amplifier (TPA) with a LiNbO$_3$ (LN) crystal has been demonstrated in this work. At 10 kHz repetition rate, a peak power of 200 W and an average power of 12 $\mu W$ have been obtained over a wide range around 2 THz; at 100 kHz repetition rate, a maximum peak power of 18 W and average power of 10.8 $\mu W$ have been obtained. The parametric gain of the LN crystal was also investigated and a modified Schwarz-Maier model was introduced to interpret the experimental results.

*doi:10.3788/COLXXXXXX.XXXXXX.*

Terahertz (THz) radiation has extensive scientific and technological applications[1-9]. In recent years, the observation of the transient phenomena in materials[8] and the real-time THz spectroscopic imaging of molecules[9] have made high repetition rate, high peak power THz radiation source an attractive research topic. The THz parametric amplifier (TPA), which is based on the stimulated polariton scattering (SPS) in some polar crystals[10], has offered the possibilities for generating high repetition rate, high power THz radiation with a broad frequency tuning range. In addition, the TPAs are operated at room-temperature and of small size and low cost, which make them affordable by most users.

The crystals used for SPS based THz source include lithium niobate (LiNbO$_3$ or LN)[11-17], lithium tantalate (LiTaO$_3$ or LT) [18], potassium titanyl phosphate (KTiOPO$_4$ or KTP)[19-23], potassium titanyl arsenate (KTiOAsO$_4$ or KTA)[24], and rubidium titanyl phosphate (RbTiOPO$_4$ or RTP)[25], among which the LN and LT crystals (always referred to as the "LN family") cover the spectral range from 0.7 to 4.7 THz. Researches on low repetition rate (10—100 Hz) LN TPA have been widely carried out [14, 15, 26]. On the contrary, high repetition rate LN TPAs are less developed. To our best knowledge, till now only one work has been published by Moriguchi et al. [13], which achieved a maximum peak / average power of 4 W / 30 µW at the repetition rate of 100 kHz. In this paper, we report our demonstration of a frequency-tunable (1.2–3.9 THz) picosecond LN TPA operated at the repetition rates of 10 kHz and 100 kHz, which achieved a maximum peak power of 200 W and 18 W, respectively. We also introduce a new model to interpret the parametric gain of the LN TPA.

The experimental setup of the LN TPA is shown in Fig. 1, which mainly comprises a picosecond pump laser, a continuous wave seed laser, and a 5% (molar) MgO-doped congruent LN crystal. The pump laser starts from a 1064.4 nm fiber laser with the output pulse duration of 14.0 ps. Following the fiber laser, a home-built neodymium-doped yttrium vanadate (Nd: YVO$_4$) regenerative amplifier boosts the pulse energy to 230 µJ / 38 µJ at 10 kHz / 100 kHz repetition rate. The seed laser is composed of an external cavity diode laser (ECDL, ECD004, moglabs) and an Yb-fiber amplifier. It has a mode-hop-free wavelength tuning range from 1068 nm to 1080 nm, over which an output power of 100 mW can be obtained.

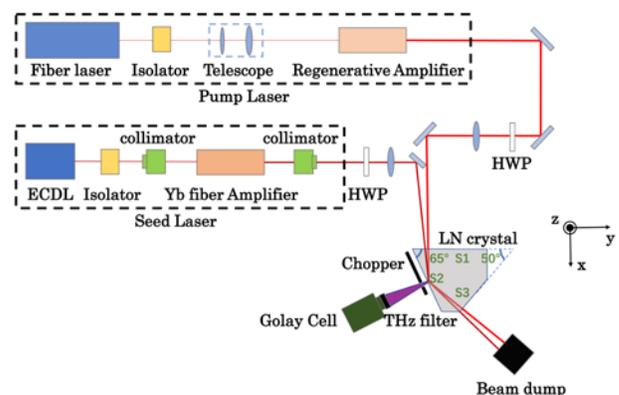

Fig. 1. A sketch of the TPA experimental setup.

The crystal is cut into a pentagon shape in x-y plane, as shown in Fig. 1. It has a thickness of 3 mm along the z axis. The pump laser propagates perpendicularly to the entrance surface (S1), where the incident Fresnel transmission coefficient is about 0.86. The seed laser has a small incident angle, which is set to meet the nonlinear phase matching condition. The polarization directions of both lasers are parallel to the optic axis of the crystal (along z axis). The intersection angle between the THz wave extraction surface (S2) and S1 is 65°. This allows a nearly perpendicular extraction of the generated THz wave without any output coupler and a total reflection of the residual pump laser and Stokes wave at S2. The crystal is mounted on a translation stage, which moves in parallel to the S2 surface. With this configuration, the distance $L$ between S1 and the THz wave extraction point on S2 (i.e., the effective parametric amplification length) can be continuously varied while the position of the extraction point keeps unchanged.

The pulse energy of the extracted THz radiation is measured by a Golay cell (GC1D, TYDEX) in combination with an optical chopper (LST202, NM Laser Products), which periodically modulates the THz radiation. In the measurements, the modulation frequency and duty cycle of the chopper were set to 7 Hz and 3%, respectively. Under this condition, the responsivity of the Golay cell (1.23 V/µJ) had been calibrated according to the procedure reported in Y. Wang's work[27] for our earlier study of an accelerator-based THz radiation source. In order to avoid the interference of the scattered light, a THz bandpass filter (LPF23.4-24, TYDEX) was inserted before the Golay cell, which has a transmittance of about 50% at 2 THz.

In our experiments, we mainly demonstrated the LN TPA at 10 kHz repetition rate. The pulse energy of the pump laser was fixed at the maximum value of 230 µJ. For a higher THz parametric gain, its radius was focused to 850 µm. This corresponds to a pump intensity of 1.5 GW/cm$^2$, which is below the laser damage threshold of the LN crystal. The seed laser power was 100 mW and the radius was similar to the pump laser. Its wavelength was set to 1071.5 nm first, corresponding to a frequency difference of 1.87 THz with respect to the pump laser. The effective amplification length for the extracted THz radiation, $L$, was chosen to be 21.2 mm, which was the maximum available value. The measured THz pulse energy was 1.2 nJ, indicating an average power of 12 µW and an extraction efficiency of about 5.2×10$^{-6}$ from the pump laser pulse.

To measure the frequency and bandwidth of the THz radiation, a scanning Fabry-Perot interferometer (TSFPI-P, TYDEX) was used in combination with the Golay cell. The results are shown in Fig. 2(a) and (b), respectively. From the interferogram in Fig. 2(a), we can observe a period of 80 µm, which indicates a frequency of 1.87 THz. This is consistent with the frequency difference between the pump laser and the Stokes wave. In the bandwidth measurement, the interferometer cavity length was scanned around 1 mm, which corresponds to a free spectral range (FSR) of 150 GHz. From the interferogram shown in Fig. 2(b) we can infer a bandwidth of FSR/2, i.e., 75 GHz. For Fourier transform limited Gaussian pulses, the THz radiation is expected to have a FWHM duration of 5.9 ps, indicating a peak power ~200 W.

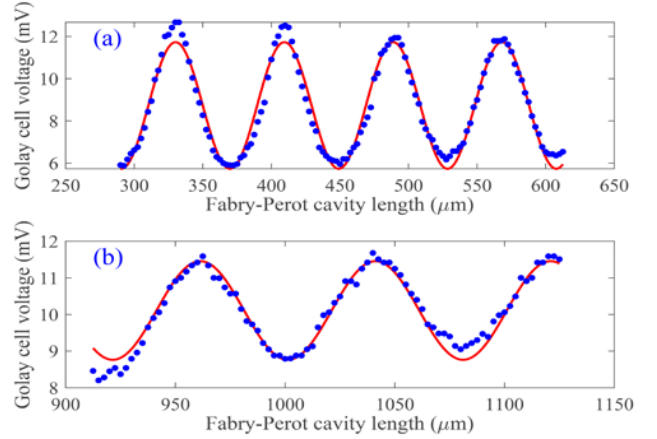

Fig. 2. Measurement result for THz radiation wavelength (a) and bandwidth (b).

We also measured the spatial distribution of the THz radiation by scanning the Golay cell (with a 1 mm diameter pinhole fixed before it) position transversely. Figure 3 presents the beam profiles at different locations downstream the S2 surface. It can be observed that the THz beam has similar horizontal (x) and vertical (y) sizes when exiting the LN crystal. However, the beam has a larger divergence in the vertical direction, and the vertical size grows much faster than the horizontal one, as shown in the figure. The THz beam is estimated to have an $M^2$ parameter of 1.3 and 2.9 in the horizontal and vertical directions, respectively.

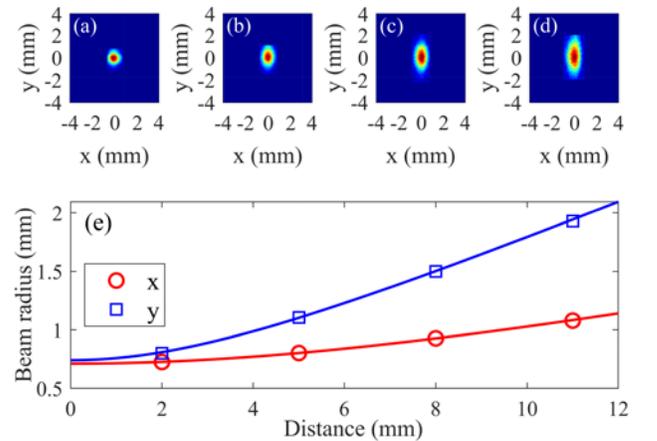

Fig. 3. THz beam profile measurement results of the LN TPA. (a-d) are THz beam profiles measured at 2.0, 5.0, 8.0, and 11.0 mm downstream the THz radiation exit surface (S2) of the LN crystal. (e) plots the beam radii for THz beam as a function of the measurement distance. The solid curves were calculated according to the estimated $M^2$ parameters.

The output THz frequency can be continuously tuned by scanning the seed laser wavelength in combination with the incident angle adjustment to meet the nonlinear phase matching condition. In our experiments, the seed laser wavelength was varied from 1068.9 nm to 1079.2 nm and its incident angle was adjusted from 1.18° to 3.95°, while the power remained unchanged at 100 mW. The achieved THz frequency tuning range was from 1.18 to 3.85 THz. Fig. 4(a) plots the output pulse energy as a function of the THz frequency for different effective parametric amplification lengths $L$. When $L = 21.2$ mm (the maximum available value), we can find a broad flat region with the highest pulse energy around 2 THz. In this case, there also exists a small peak around 3.6 THz, which is due to the phonon-polariton damping of low frequency modes in the LN crystal[29, 30].

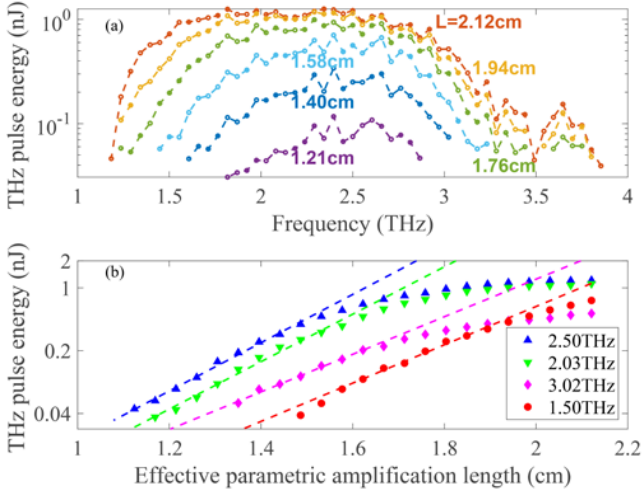

Fig. 4. (a) Measured THz pulse energy as a function of frequency for different effective parametric amplification lengths $L$. (b) Measured pulse energy THz wave output as a function of $L$ at four different THz frequencies.

In Fig. 4(b), we plotted the THz pulse energy at four different frequencies as a function of $L$. To derive the parametric gain $g_{THz}$, the data points within the exponential growth region were fitted according to

$$E_{pulse} = E_0 e^{g_{THz}L}, \quad (2)$$

where $E_{pulse}$ is the pulse energy of THz radiation and $E_0$ its initial value. The fitted curves were included in Fig. 4(b), while a complete set of measured $g_{THz}$ was plotted in Fig. 5 as a function of THz frequency. As can be seen from the figure, a peak value of $g_{THz}$ occurs at 2.39 THz.

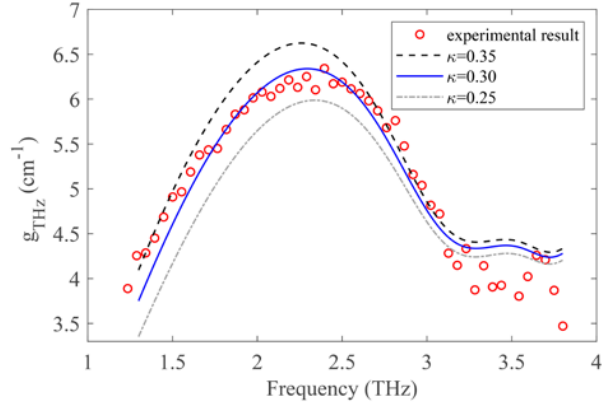

Fig. 5. THz parametric gain ($g_{THz}$) vs frequency. The curves represent theoretical results calculated using Eq. (3), for which 3 different $\kappa$ values were used.

To interpret the experimental results, we introduce a modified Schwarz-Maier model for the LN crystal. In this model, $g_{THz}$ is given by

$$g_{THz} = \frac{\alpha_T + \alpha_{pr}}{2}\left[\sqrt{1 + 16\cos\varphi\left(\frac{\Gamma g_0}{\alpha_T + \alpha_{pr}}\right)^2} - 1\right], \quad (3)$$

where $\alpha_T$ is the absorption coefficient of THz wave in the crystal, $\varphi$ is the phase matching angle between the pump laser and the THz wave, and $\alpha_{pr} = \sin\varphi/\kappa w$ accounts for the walk-off loss between the two beams[29] with $w$ the pump beam radius, and $\kappa$ the ratio between the effective interaction region and $w$. The parametric gain under low-loss limit, $g_0$, can be calculated as

$$g_0 = \sqrt{\frac{\pi\omega_s\omega_T I_p}{2c^3 n_s n_p n_T}}\chi_p, \quad (4)$$

where $I_p$ is the intensity of the pump laser, $\omega_s$ and $\omega_T$ represent the angular frequency of the Stokes wave and THz wave, respectively, $n_p$, $n_s$, and $n_T$ represent the refractive index of the pump laser, Stokes wave and THz wave in LN crystals, respectively, and $\chi_p$ is the effective nonlinear coefficient with electronic and ionic contributions taken into account. $\Gamma = \sqrt{w/(w+2/\alpha_T)}$ is newly introduced referring to some earlier works[31, 32], accounting for the mode area mismatch between the pump laser and THz beam, which has not been considered when calculating $g_0$.

In Eq. (3), $\alpha_T$, $\varphi$, $g_0$, and $\Gamma$ can be obtained directly or derived based on the LN crystal data[33], the pump laser radius and pulse energy, the Stokes wave frequency, and the THz wave frequency. The remaining parameter $\alpha_{pr}$ has a dependency on $\kappa$, which cannot be determined straightforwardly. We therefore scanned $\kappa$ to find the best matching between the equation and experimental data. The optimized $\kappa$ was found to be 0.30 in this case. The solid curve in Fig. 5 represents the THz parametric gain calculated as a function of THz frequency, which agrees well with the experimental result.

We finally demonstrated the LN TPA at 100 kHz repetition rate. In the experiments, the pump laser pulse energy was 38 μJ. The measured THz pulse energy was 0.11 nJ at 2 THz, corresponding to an extraction efficiency of $2.9 \times 10^{-6}$. The peak power of THz radiation was estimated to be 18 W.

In conclusion, we have demonstrated a high repetition rate, narrowband, picosecond TPA with a LN crystal. At 10 kHz repetition rate, a peak power of 200 W and an average power of 12 μW have been obtained over a wide range around 2 THz. At 100 kHz repetition rate, a maximum peak power of 18 W and average power of 10.8 μW have been obtained. We have also investigated the parametric gain of the LN TPA. Especially, we introduced an analytical formula for THz gain, which interpreted the experimental results quite well.

The authors thank Yen-Chieh Huang of National Tsinghua University (Hsinchu) for his valuable help on developing the laser-based THz source. This work is supported by National Key Research and Development Program of China (Grant No. 2017YFA0701000) and National Natural Science Foundation of China (Grant No. 11735002).